\documentclass[aps,prl,twocolumn,floatfix,numerical]{revtex4-1}

\usepackage{graphicx}
\usepackage{epstopdf} 
\usepackage{dcolumn}
\usepackage{bm}
\usepackage{color}

\begin{document}

%
%

\title{Three dimensional density cavities in guide field collisionless magnetic reconnection}

\author{S. Markidis}
\email{markidis@pdc.kth.se}
\affiliation{Centrum voor Plasma-Astrofysica, Departement Wiskunde, Katholieke Universiteit Leuven, Leuven, Belgium}
\affiliation{PDC Center for High Performance Computing, KTH Royal Institute of Technology, Stockholm, Sweden}

\author{G. Lapenta}
\author{A. Divin}
\affiliation{Centrum voor Plasma-Astrofysica,
Departement Wiskunde, Katholieke Universiteit Leuven,  Leuven, Belgium}
\author{M. Goldman}
\author{D. Newman}
\affiliation{Department of Physics and CIPS, University of Colorado, Boulder, Colorado, USA}
\author{L. Andersson}
\affiliation{Laboratory for Atmospheric and Space Physics, University of Colorado, Boulder, Colorado, USA}

\date{\today}

\begin{abstract}
Particle-in-Cell simulations of collisionless magnetic reconnection with a guide field reveal for the first time the three dimensional features of the low density regions along the magnetic reconnection separatrices, the so-called {\em cavities}. It is found that structures with further lower density develop within the cavities. Because their appearance is similar to the rib shape, these formations are here called {\em low density ribs}. Their location remains approximately fixed in time and their density progressively decreases, as electron currents along the cavities evacuate them. They develop along the magnetic field lines and are supported by a strong perpendicular electric field that oscillates in space. In addition, bipolar parallel electric field structures form as isolated spheres between the cavities and the outflow plasma, along the direction of the low density ribs and of magnetic field lines. 
\end{abstract}
\maketitle

%
%

\section{Introduction}
Magnetic reconnection is the topology reorganization of disconnected magnetic regions into a new configuration with concurrent conversion of magnetic energy into plasma kinetic energy. The magnetic energy is released in form of plasma heating and acceleration: plasma (inflow plasma) drifts toward the region where magnetic reconnection takes place and expelled from it forming the reconnection jets (outflow plasma). The surfaces between the new magnetic regions, dividing the inflow and outflow plasmas are called separatrices.  Separatrices are thin boundary layers between plasmas with different properties (density, temperature, and flow pattern)~\cite{Khotyaintsev:2006,Retino:2006,Fujimoto:2011} and subject to instabilities, wave activity and turbulence~\cite{Daughton:2011, Andrey:2011, Pritchett:2004}. One of the distinctive features of collisionless magnetic reconnection is the formation of localized low density regions along the separatrices~\cite{Lu:2010,Zhou:2011}. These depletion areas are called {\em cavities}. 

One of the most common and simple reconnection configurations is that of a sheared magnetic field layer, with a superimposed uniform magnetic field (called {\em guide field}) perpendicular to the plane of the reversing magnetic field (the reconnection plane). The configuration provides a simple setup, but still realistic enough to compare simulation results with satellite observations of magnetopause~\cite{birn-book,Yamada:2010,Paschmann:2008} and magnetotail~\cite{Eastwood:2010}. In addition, this configuration results a useful approximation of magnetic field lines in nuclear fusion devices, such as the case of the tokamaks and reversed field pinch machines, where a toroidal field is present. In presence of guide field, separatrices display anti-symmetric features with the respect to the X point: the cavities develop only along two separatrices, while regions of increased density form along the other two separatrices~\cite{Pritchett:2004, Swisdak:2005, Lapenta:2010}. In addition, strong electron beams move toward the X line along the cavities separatrices. Streaming instabilities, such as the Buneman, two-stream, and lower hybrid instabilities are triggered by electron beams in the cavity regions. As macroscopic signature of these instabilities, bipolar electric field structures are detected in simulations~\cite{Drake:2003, Cattell:2005, Goldman:2008}, magnetospheric observations~\cite{andersson,Khotyaintsev:2010} and laboratory experiments~\cite{Fox:2008}.

The cavities were first  identified in simulations by {\em Shay et al.}~\cite{Shay:2001}, as one of the characteristic signatures of Hall magnetic reconnection. The decoupling of ions and electrons leads to an electron current system surrounding the magnetic separatrices, as suggested by the theory of Hall reconnection~\cite{Uzdensky:2006, Korovinskiy:2008}. The net parallel electric field strongly accelerates the electrons in this region, thus reducing the local density~\cite{Drake:2005}.

Previous studies focused on analyzing the structure and scaling properties of the cavities in two dimensional Particle-in-Cell simulations. It was found that cavities size scales as the ion skin depth, and their density is insensitive to the simulation mass ratio~\cite{Lapenta:2010}. In addition the simulation results were confirmed by direct observations of cavities in the magnetosphere by the Cluster spacecrafts~\cite{Lu:2010,Zhou:2011}.  The majority of previous studies mainly regarded to the two dimensional structures of the cavities, while the three dimensional structure was not studied in detail. This work focuses on the study of magnetic reconnection separatrices and three-dimensional cavity structure, presenting two main results. First, it reveals the three-dimensional structures of the cavities. They appear as low density layers. Regions, characterized by further lower density, here called {\em low density ribs} form within the cavities. They develop along the magnetic field lines and are supported by strong perpendicular electric fields. Second, it shows that bipolar electric field signatures are present as spherical formations. They are aligned with the low density ribs and magnetic field lines, and they are located between the cavity and the outflow plasma.

This paper is organized as follows. The first section describes the simulation set-up and the parameters in use. The second section presents the evolution of three dimensional guide field magnetic reconnection, the cavity and low density ribs formation, and the presence of bipolar parallel electric field structures in their proximity. The third section discusses the results and the possible origins of three-dimensional structure of the cavities. Finally, the conclusion section summarizes the results.

\section{Simulation Set-up}
Simulations are carried out in a three-dimensional system, where an Harris current sheet equilibrium is initially imposed~\cite{harris}. The $z$ coordinate is taken along the Harris sheet current, while the $x$-$y$ plane is the reconnection plane. The $x$, $y$ and $z$ simulation coordinates correspond to the $-x_{GSM}$, $z_{GSM}$, and $-y_{GSM}$ coordinates in the Geocentric Solar Magnetospheric (GSM) system and to Earth-Sun, North-South, and dawn-dusk directions in Earth magnetotail.

The plasma density profile is initialized as:  
\begin{equation}
n(y) = n_0 \cosh^{-2}\left(\frac{y - L_y/2}{\lambda}\right)  + n_{b}
\end{equation}
The peak density $n_0$ is the reference density, while the background density $n_{b} = 0.1\ n_0$; $L_x \times L_y  \times L_z = 20 \ d_i \times 15 \ d_i \times 10 \ d_i$ are the simulation box lengths and $\lambda = 0.5 \ d_i$ is the half-width of the current sheet. The ion inertial length is $d_i = c/\omega_{pi}$,  with $c$ the speed of light in vacuum, $\omega_{pi} = \sqrt{4 \pi n_0e^2/m_i}$ the ion plasma frequency, $e$ the elementary charge, and $m_i$ the ion mass. The  electron mass is $m_e = m_i/256$.

A magnetic field in the $x$ direction is initialized to satisfy the Harris equilibrium force balance:
 \begin{equation}
B_x(y)=B_0\tanh\left(\frac{y - L_y/2}{\lambda}\right).
\end{equation}
A uniform guide field $B_g = B_0$ along the $z$ direction is added to the magnetic field profile and a perturbation of the $z$ component of the vector potential, $\delta A_z$, is applied to initiate magnetic reconnection:
 \begin{equation}\label{perturb}
\delta A_z=A_{z0} \cos(2\pi x/L_\Delta)cos(\pi y/L_\Delta)e^{-(x^2+y^2)/\lambda^2},
 \end{equation}
with $L_\Delta=10 \ \lambda$.

The particles are initialized with a Maxwellian velocity distribution. The electron thermal velocity $v_{the}/c = (T_e/(m_e c^2))^{1/2} =  0.045$, while the ion temperature $T_i = 5 \ T_e$. The simulation time step is $\omega_{pi}\Delta t = \ 0.125$. In this simulation set-up, $\omega_{pi}/\Omega_{ci} = c/ V_A = 103$, where $\Omega_{ci} = e B_0/m_i$ is the ion cyclofrequency and $V_{A}$ is the Alfv\'en velocity calculated with $B= B_0$ and $n = n_0$. The ion Larmor gyro-radius is $0.65 \ d_i$. 

The grid is composed of $256 \times 192 \times 128$ cells, resulting in a grid spacing $\Delta x  = \Delta y = \Delta z = 0.08 \ d_i$. In total $3\times 10^9$ computational particles are in use. The boundaries are periodic in the $x$ and $z$ directions, while are perfect conductor and reflecting boundary conditions for fields and particles respectively in the $y$ direction. Simulations are carried out with the parallel implicit Particle-in-Cell {\em iPIC3D} code~\cite{Markidis:2010}.

\section{Results}
The evolution of magnetic reconnection is well depicted by the history of the electron current sheet, shown in a volume plot at different times in Figure~1. Different colors represent the intensity of electron current density: the green and blue colors identify high and low current density areas respectively. The initial thin current sheet breaks in the region where the magnetic reconnection was triggered by the initial perturbation, and separates in two parts moving out mainly along the $x$ direction. It is important to note that the presence of the guide filed alters the direction of the reconnection jets: their directions are not purely in the $x$ direction, as in the case of antiparallel reconnection, but they are tilted by the guide field~\cite{Eastwood:2010}. As result of the reconnection dynamics, thin current surfaces, the separatrices, develop between the inflow and outflow plasmas around time $\Omega_{ci} t = 12.1$. Finally, the separatrices become unstable at later times with the appearance of filamentary structures along them.
\begin{figure}[ht]
\includegraphics[width=\columnwidth]{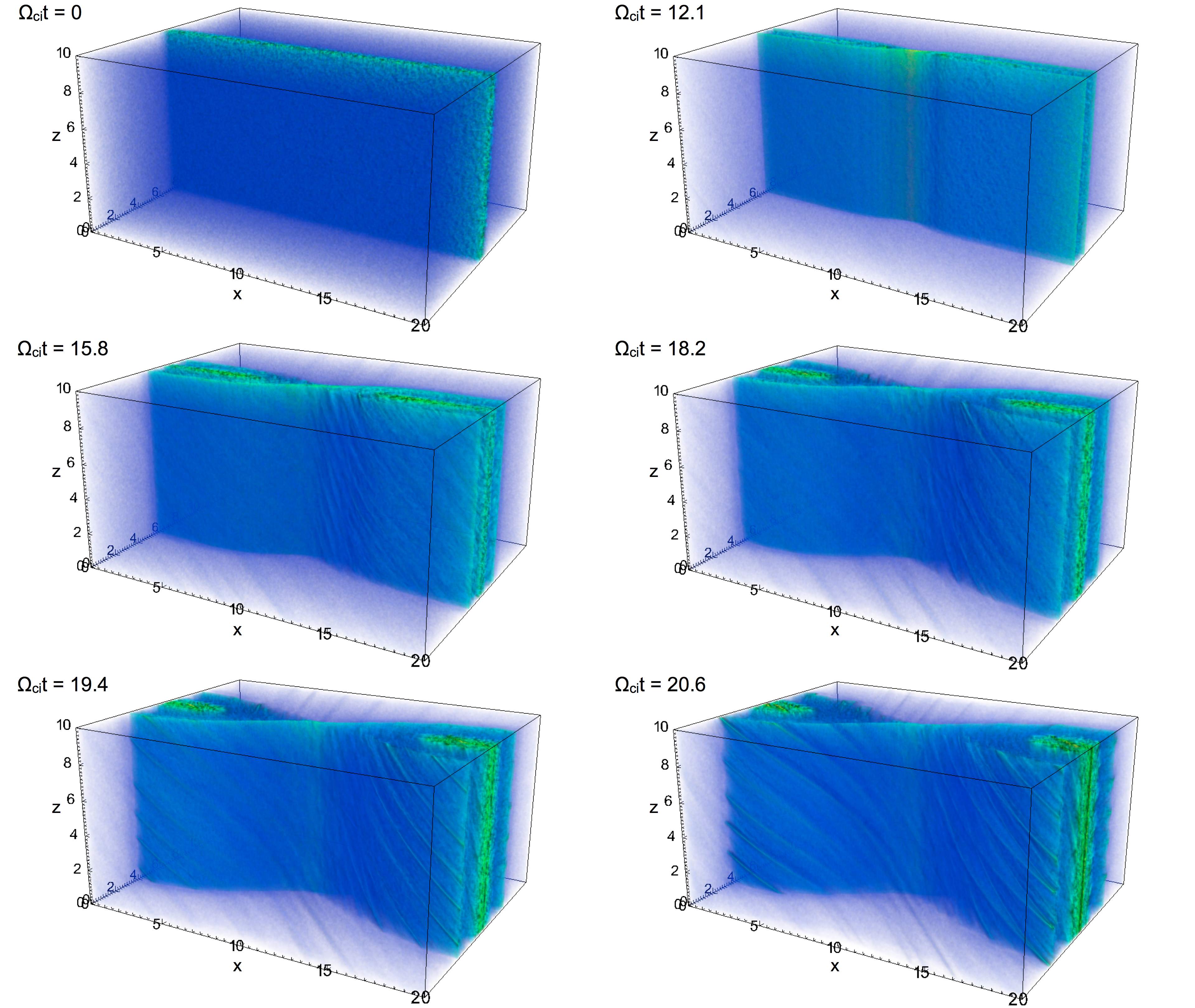}
\caption{Volume plot of the electron current density at different times. The initial Harris current sheet breaks in two reconnection jets, moving away from the $x$ line. The electron current develops filamentary structures along the separatrices at later times.}
\label{volumePlotJe}
\end{figure}

A new magnetic field topology is created from two distinct initial magnetic regions as a result of reconnection. Figure~2 shows first a snapshot of the magnetic field lines at time $\Omega_{ci} t = 12.1$ in panel a, and then an isosurface plot of the reconnection electric field $E_z$  from two different points of view in panels b and c. The reconnection electric field $E_z$ reaches the peak value of $0.35 \ B_0 V_A/c$ in a cylindrical region around the X line. The value of the reconnection electric field is in good agreement with the value obtained in two dimensional simulations of kinetic guide field reconnection~\cite{Birn:2001, Markidis:2011,Lapenta:2011}.  
\begin{figure}[ht]
\includegraphics[width=0.68 \columnwidth]{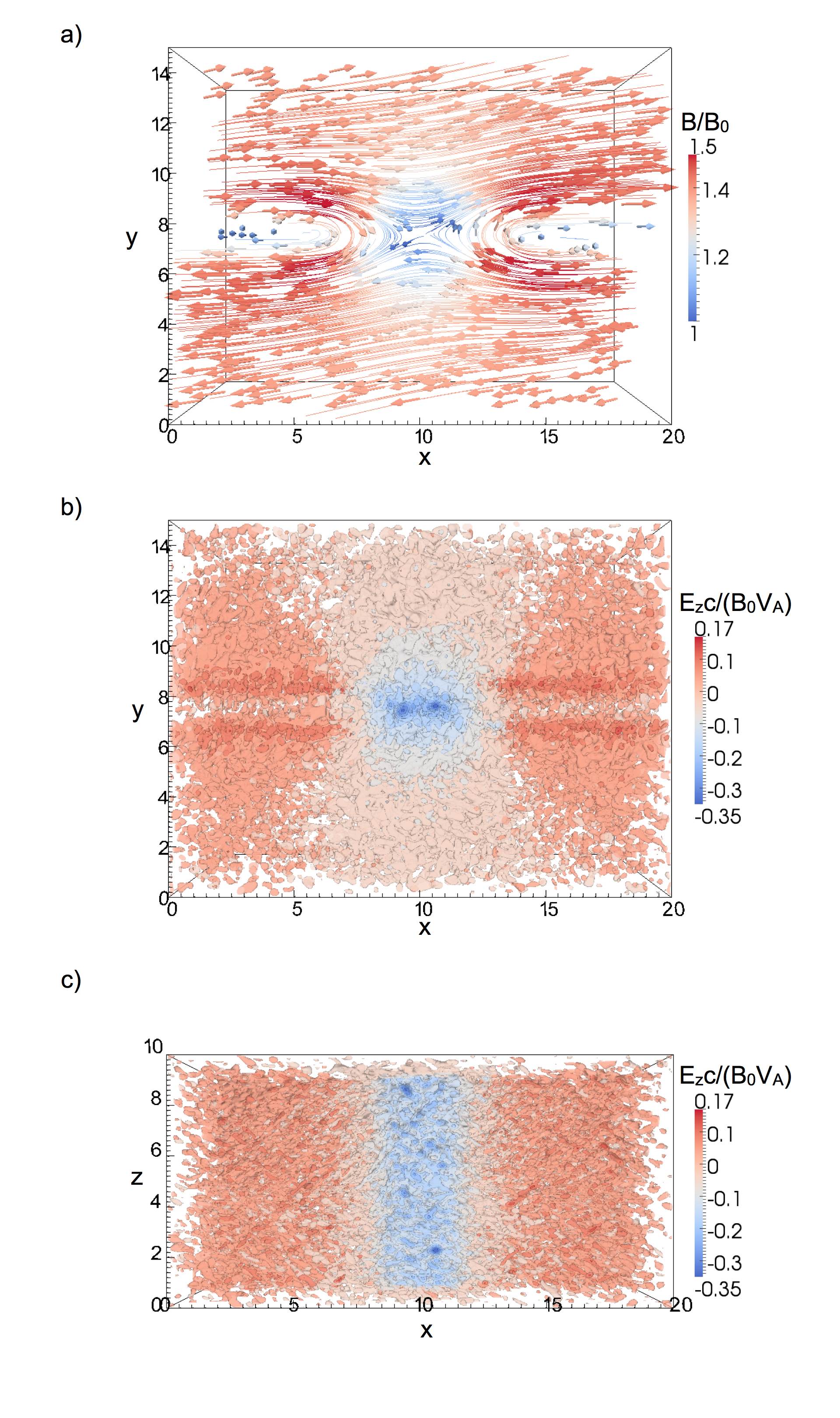}
\caption{Magnetic field lines at time $\Omega_{ci} t = 12.1$ in panel a. The color represents the magnitude of the magnetic field, normalized to the guide magnetic field $B_0$. Panels b and c show the reconnection electric field, $E_z$, normalized to $B_0 V_A/c$ from two different points of view at time $\Omega_{ci} t = 12.1$.}
\label{Blines}
\end{figure}

As magnetic reconnection evolves, areas of decreased density regions, the {\em cavities}, develop along separatrices. Figure~3 shows a plot of the electron density evolution in an area around the X line ($x= 4.8 \ d_i - 14.5 \ d_i, y = 4 \ d_i - 11.2 \ d_i$) on the reconnection plane at $z = L_z/2$. The electron density is normalized to the peak current sheet density $n_0$, and the maximum value of the color-bar is fixed to $0.3 \ n_0$ to investigate the low density regions. The typical density patter found in other two dimensional Particle-in-Cell simulations is recovered. The density configuration is asymmetric with respect to the X point. The top left and bottom right quadrants of the four panels show the cavities, the low density regions in blue color in the contourplot.
\begin{figure}[ht]
\includegraphics[width=\columnwidth]{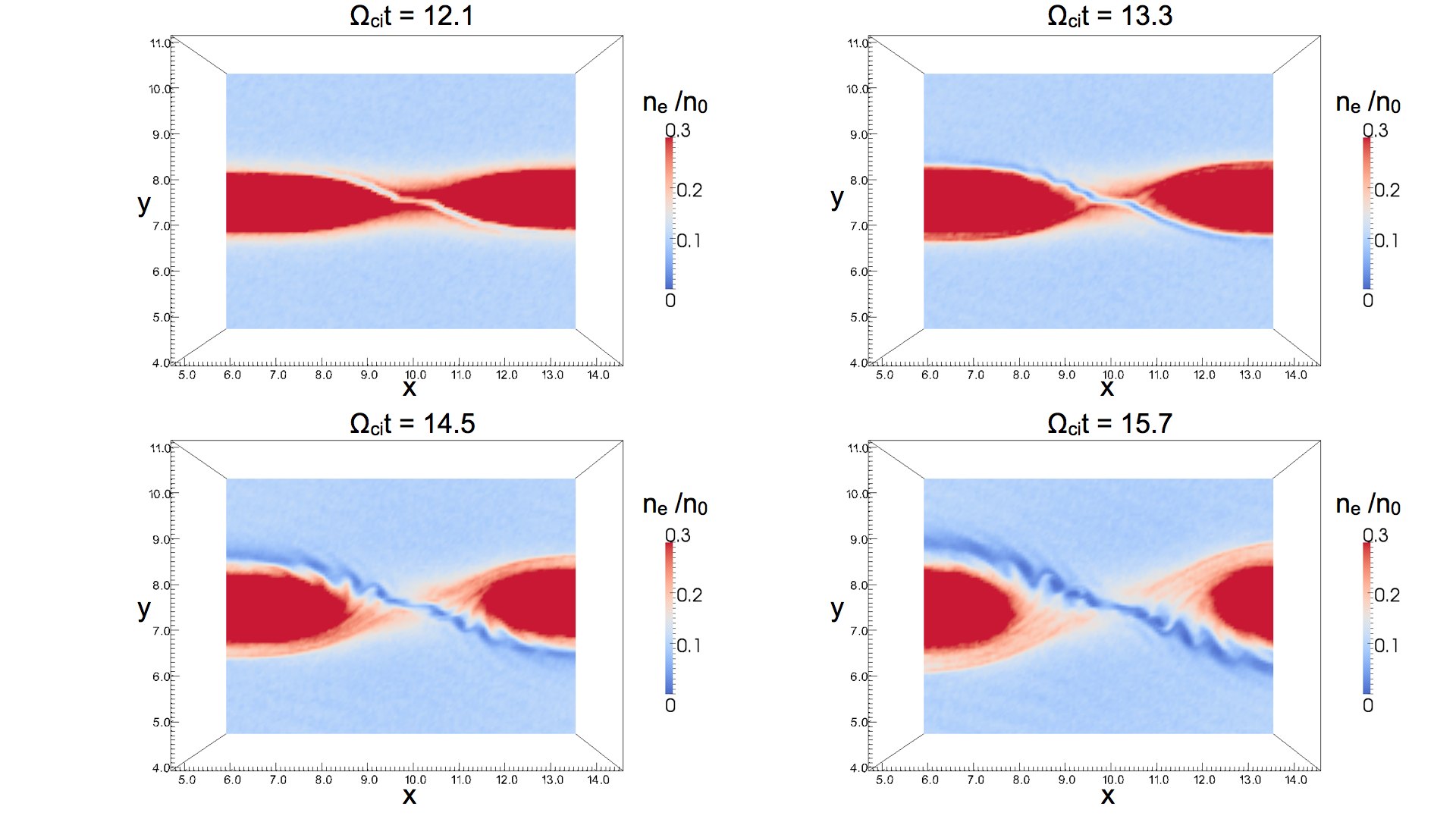}
\caption{Contour plots of the electron density in the region $x= 4.8 \ d_i - 14.5 \ d_i, y = 4 \ d_i - 11.2 \ d_i, z = L_z/2$ at different times. The plot shows the formation of the density cavity layers (blue regions), and the successive development of density ripples, reminiscent of Kelvin-Helmholtz vortices, along the cavities. The local magnetic field in the cavity is approximately $45^\circ$ out of the presented plane.}
\end{figure}
The cavities form at time $\Omega_{ci} t = 12.1$, they are perturbed by a wave structure at time $\Omega_{ci} t = 13.3$ and are strongly distorted after time $\Omega_{ci} t = 14.5$. Small spatial scale low density ripples develop along the cavities. These structures are similar in shape to the Kelvin-Helmholtz vortices \cite{Andrey:2011}. Note that in the density cavities of Figure~3 the local magnetic field is approximately $45^\circ$ out of the presented plane.

\subsection{Low density ribs within the cavities}
The three dimensional structure of the cavities is studied in detail in this section. The cavities appeared as thin electron current layers in two dimensional Particle-in-Cell simulations and in the contourplot on the plane $z = L_z/2$ in Figure~3. Figure~ 4 panel a, shows two isosurface plots of the electron density at time $\Omega_{ci} t = 14.5$. The electron density is normalized to the background density $n_b = 0.1 \ n_0$. The isosurface $n_e = 0.65 \ n_b$ (orange color) shows the cavity layer, approximately $1\ d_i$ thick; the isosurface $n_e = 0.4 \ n_b$ (grey color) reveals that regions with further lower density are embedded in the cavities. These structures are similar in shape to ribs, and for this reason we call these regions {\em low density ribs}. Separate low density ribs show nearly identical features. They have approximately equal size and bend along the magnetic field lines directions, close to the reconnection X line. 

Figure~4 panel b shows the density isosurfaces for decreasing electron density values at successive times:  the grey, red, blue and green isosurfaces represent regions with $n_e = 0.45 \ n_b$ at time $\Omega_{ci} t = 12.1$, $n_e = 0.3\  n_b$ at time $\Omega_{ci} t = 13.3$, $n_e = 0.16 \ n_b$ at time $\Omega_{ci} t = 14.5$ and $n_e = 0.15\  n_b$ at time $\Omega_{ci} t = 15.7$ respectively. Two main results are found analyzing Figure~4 panel b. 
First, the cavities are perturbed by density waves that are almost stationary in space, far from the X line ($ 12\ d_i< x < 15 \ d_i$) in a time period $12.1 <\Omega_{ci} t < 15.7$. Closer to the X line ($ 10 \ d_i < x < 12  \ d_i$), the density waves start to overlap at different times, probably because of reconnection non-stationary dynamics and of the temporal variation of the electron velocity near the X line. Second, the density in the low density ribs decreases in time. Note that the electron population can develop areas with density as low as $n_e = 0.2 \ n_b$ when the ambient background plasma starts to dominate the separatrix edge.

\begin{figure}[ht]
\includegraphics[width=0.8 \columnwidth]{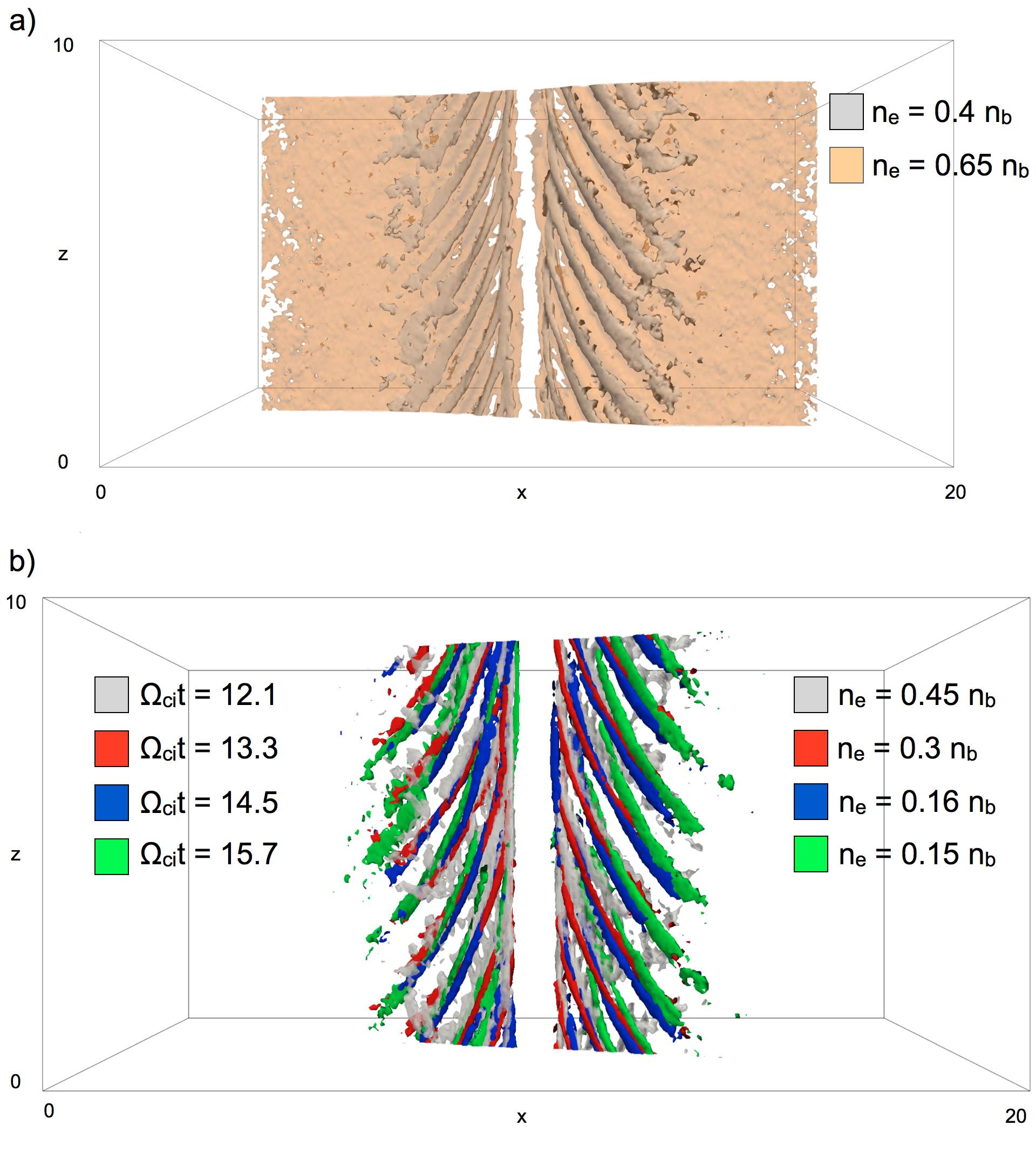}
\caption{Panel a shows the two electron density isosurface plots for $n = 0.65 \ n_b$ (orange color) and $n = 0.4 \ n_b$ (grey color) at time $\Omega_{ci} t = 14.5$, revealing the cavities and the low density ribs within them. Panel b shows the electron density isosurface plots at successive times ($\Omega_{ci} t = 12.1$ in grey, $\Omega_{ci} t = 13.3$ in red, $\Omega_{ci} t = 14.5$ in blue, and $\Omega_{ci} t = 15.7$ in green) for decreasing electron densities ($n_e = 0.45 \ n_b$ in grey, $n_e = 0.3 \ n_b$ in red, $n_e = 0.16 \ n_b$ in blue, and $n_e = 0.12 \ n_b$ in green). The location of the low density ribs is approximately fixed in time, and their density progressively decreases.}
\end{figure}

An intense parallel electron current develops along the cavities and contributes to the formation of the low density ribs by progressively decreasing the density in localized channels. This mechanism was identified previously using two dimensional Particle-in-Cell simulations in Refs.~\cite{Lu:2010,Zhou:2011}, and our study confirms that electron acceleration inside the cavities is the main mechanism for creating the cavity density depletion.
Figure~5 shows electron current streamlines in the cavity regions at time $\Omega_{ci} t = 14.5$. The electron current streamlines are organized in bundles directed along the cavity channels. The electron velocity reaches approximately the peak $ 20 \ V_{Ae}$, where $V_{Ae}$ is the electron Alfv\'en velocity.
\begin{figure}[ht]
\includegraphics[width=0.6 \columnwidth]{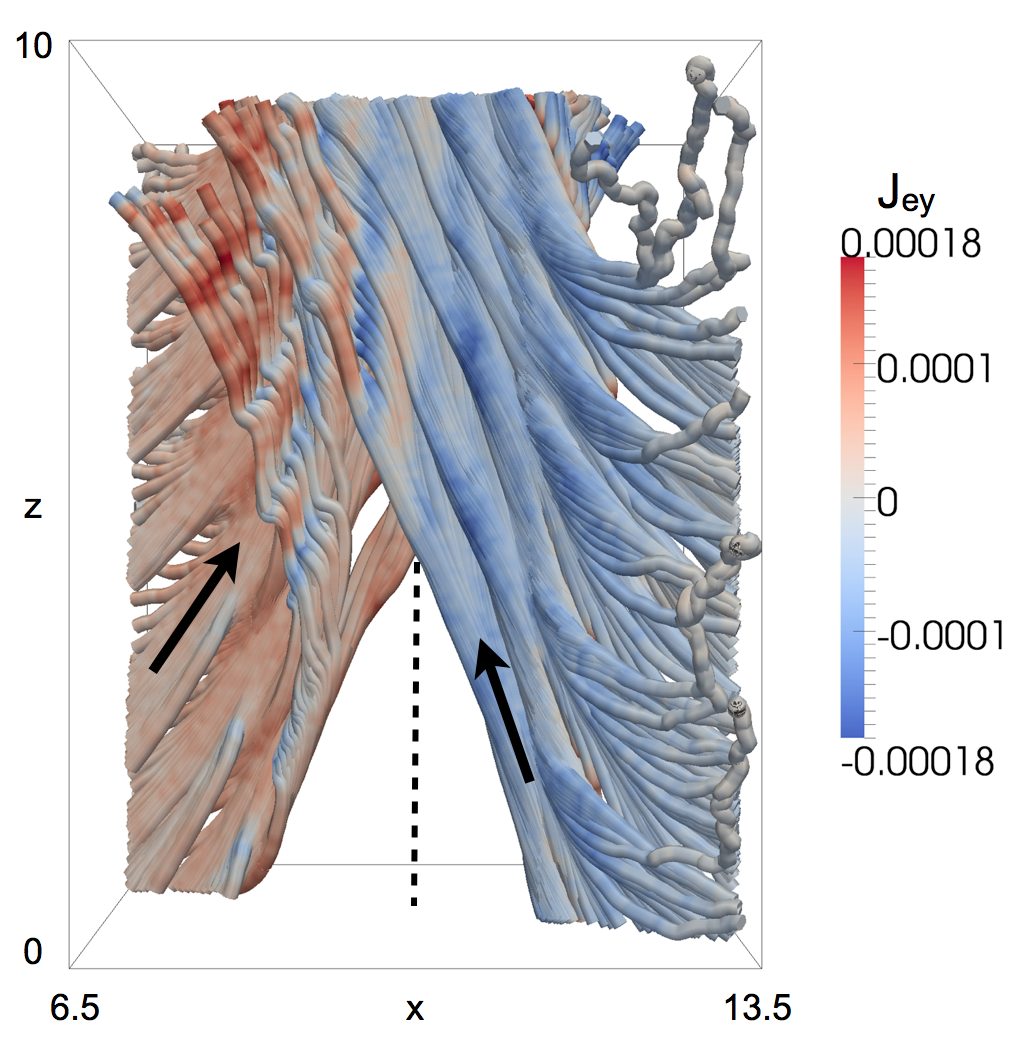}
\caption{Electron current streamlines in the cavity region at time $\Omega_{ci} t = 14.5$. The color indicates the $J_y$ value, normalized to $e n_0 c$. The electron current streamlines are organized in bundle in the cavity region. The X line is along the $z$ direction at $x=10 \ d_i, y=7.5 \ d_i$ (dashed black line). The streamlines at $x < 10 \ d_i$ (left side in the $x$ direction) and at $x > 10 \ d_i$ (right side in the $x$ direction) originate from seed lines at$x=6.5 \ d_i, y=8.5 \ d_i$ and $x=13.5 \ d_i, y = 6.5 \ d_i$. The arrows indicate the electron flow direction in the $x-z$ plane far from the X line.}
\end{figure}

The low density ribs are aligned with magnetic field lines. The two dimensional studies performed in Ref.~\cite{Andrey:2011} reveal that the unstable electron Kelvin-Helmholtz waves grow rapidly in a plane perpendicular to the local magnetic field. Figure~6 shows the magnetic field lines (red tubes): they are parallel to the grey $n_e = 0.36 \ n_b$ isosurface, which identifies the low density ribs, at time $\Omega_{ci} t = 14.5$. The low density rib structures are supported by the strong electric fields. An isosurface plot of the perpendicular electric field intensity ($\mathbf{E}_{\perp} = \mathbf{E} - \mathbf{E} \cdot \mathbf{B}/ |\mathbf{B}|$), shows the presence of a perpendicular electric field that oscillates in space in proximity of the cavities. The perpendicular electric field oscillates in space between $ \approx -1.5 \ B_0V_A/c$ and $\approx 1.0 \ B_0V_A/c$ values, with an oscillation wavelength around one Larmor gyro-radius. Its maximum intensity is approximately 4.5 times the reconnection electric field $E_z$ (Figure~2 panels b and c).  

\begin{figure}[ht]
\includegraphics[width=\columnwidth]{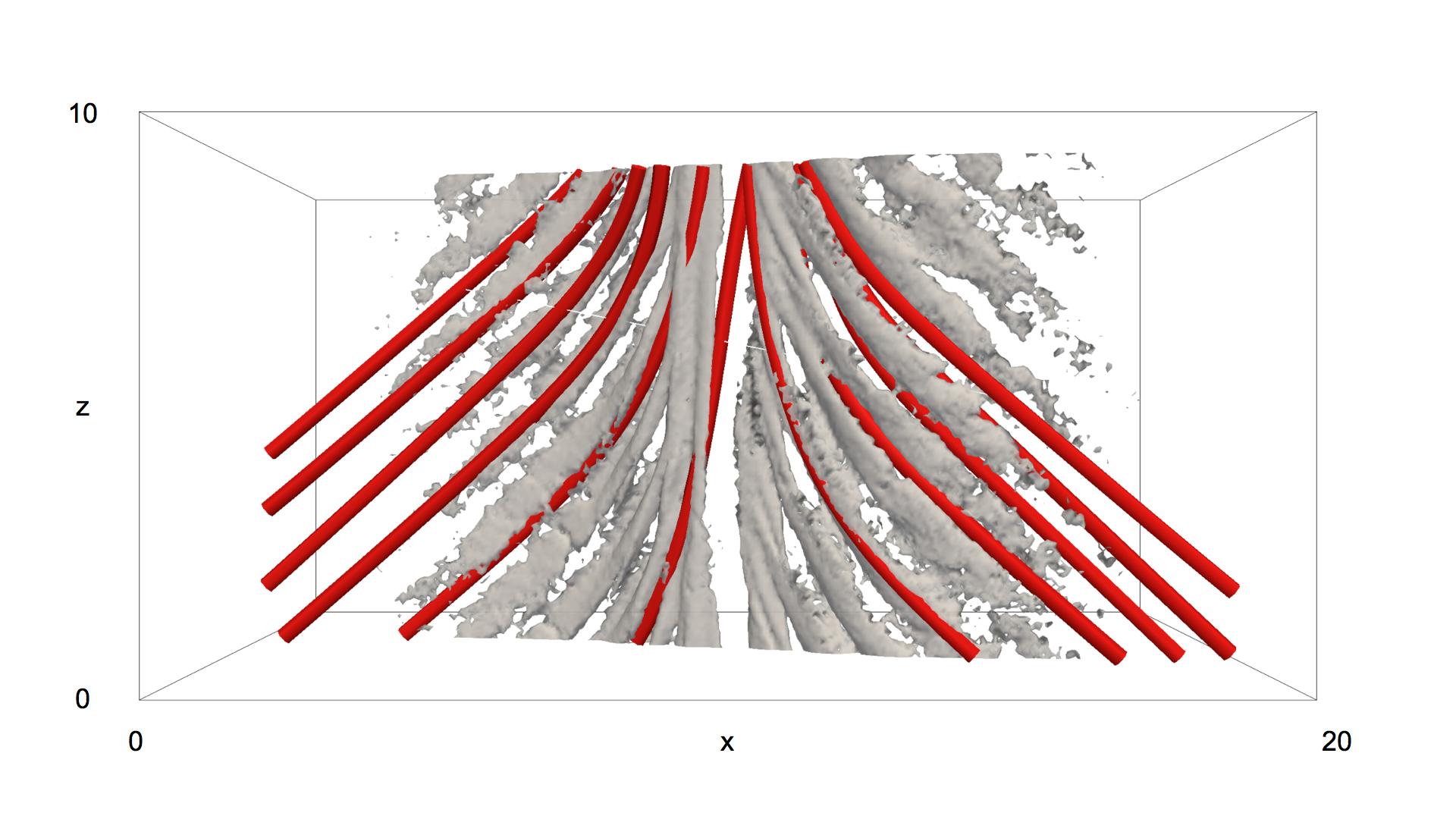}
\caption{Electron density isosurface plot for $n_e = 0.36 \ n_b$ with magnetic field lines in red tubes at time $\Omega_{ci} t = 14.5$. The low density ribs are aligned with the magnetic field lines.}
\end{figure}
\begin{figure}[ht]
\includegraphics[width=\columnwidth]{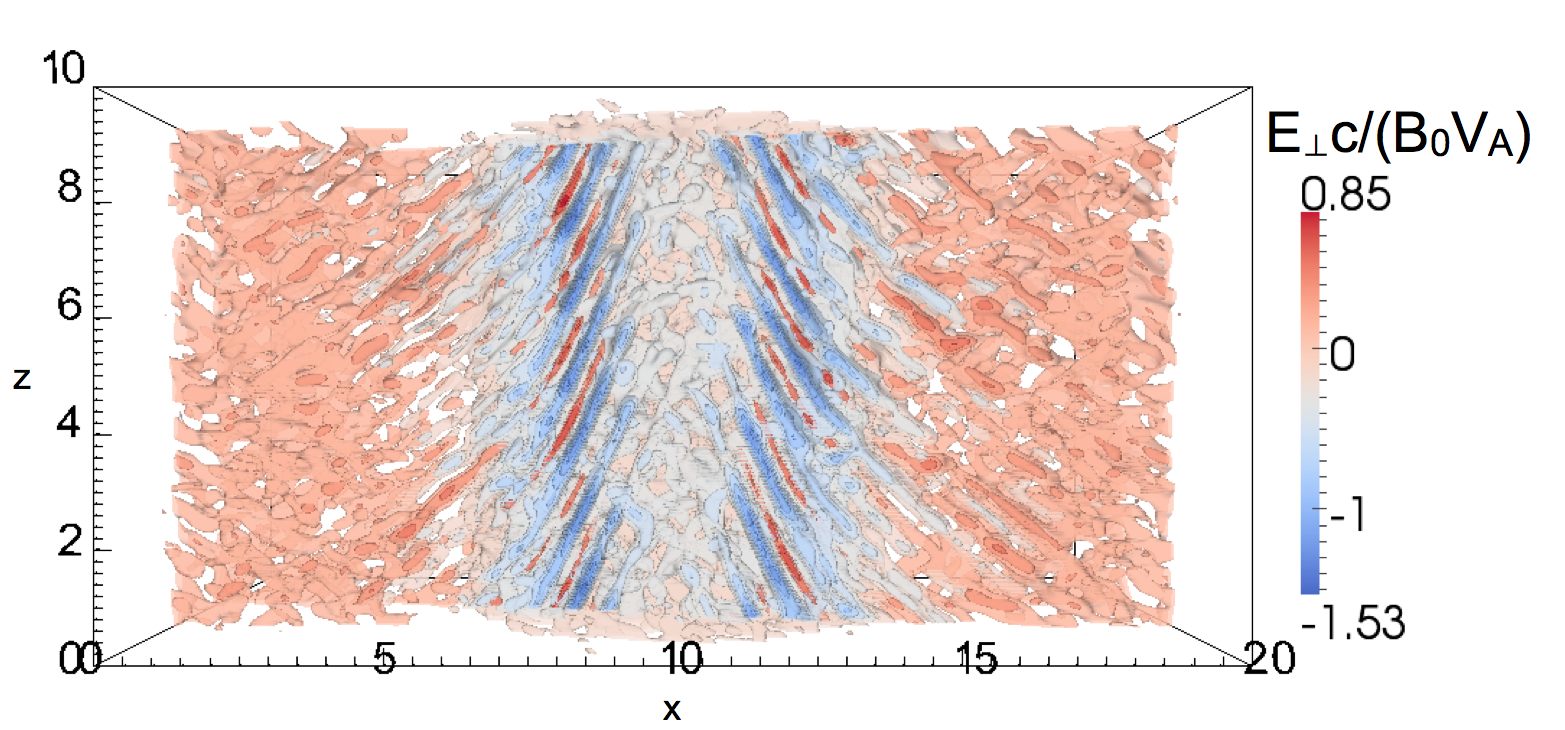}
\caption{Perpendicular electric field intensity isosurface plot at time $\Omega_{ci} t = 14.5$. The perpendicular electric field reaches high intensity values (approximately 4.5 the reconnection electric field) and it oscillates in space supporting the low density rib structures.}
\end{figure}




\subsection{Bipolar electric field structures }
The carried-out simulations confirm the presence of bipolar parallel electric field ($E_{//} =  \mathbf{E} \cdot \mathbf{B}/|\mathbf{B}|$) signatures along the cavities, reported by previous simulation studies~\cite{Drake:2003, Che:2009,Lapenta:2011}. Figure~8 shows an isosurface plot of $E_{//} = \pm 0.3 \ B_0V_A/c $ with density isosurfaces $n_e = 0.4 \ n_b, 0.65 \ n_b$ from two different points of view. Note that the net parallel electric field, that determines the electron acceleration inside the cavities, is much weaker than the bipolar electric fields and it does not appear in the plot. The presence of  bipolar parallel electric field structures along the cavities, appearing as red and blue hemispheres, is evident. The plot reveals that the bipolar structures are isolated spherical formations. The peak intensity of the bipolar parallel electric field signatures is of order $0.3 \ B_0V_A/c$.  
The combination of electron density and parallel electric field plots in Figure~8 shows that the bipolar electric field structures are elongated in the direction of the low density ribs, and of the magnetic field lines. Solitary electrostatic waves tend to appear between consecutive low density ribs in relatively higher density regions of the cavities. The positive hemisphere of the bipolar electric field precedes the negative part in the direction toward the X point along the separatrix. Panel b reveals that the the bipolar electric field structures form between the cavities and the outflow plasmas.

\begin{figure}[ht]
\includegraphics[width=\columnwidth]{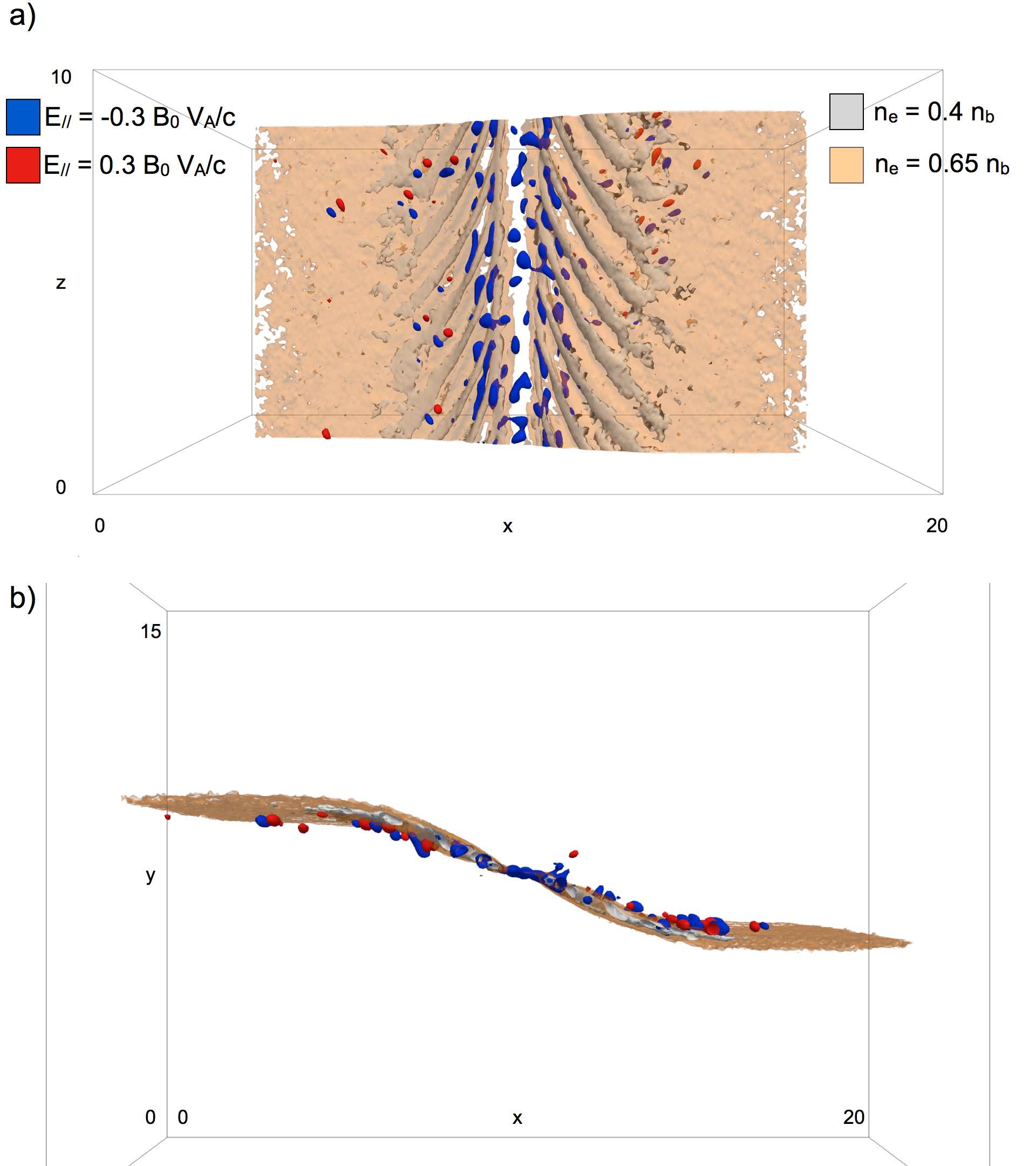}
\caption{ Electron density isosurface plot for $n = 0.4 \ n_b, 0.65 \ n_b$ in orange and grey colors with parallel electric field isosurface plot for $E_{//} = \pm 0.3 \ B_0V_A/c$ in red and blue colors at time $\Omega_{ci} t = 14.5$ from two different points of view. The bipolar parallel electric field  structures develop along the low density ribs and magnetic field lines direction between the cavities and the outflow regions. Supporting material is provided online to analyze this configuration from different points of view. }
\label{EparEperp}
\end{figure}


\section{Discussion}
The reported Particle-in-Cell simulation shows the formation of the cavity layers and of low density ribs within them along the reconnection separatrices. The low density ribs develop along the magnetic field lines, and their location remain approximately fixed in time. Their density decreases progressively in time as electron currents evacuate them. A strong perpendicular electric field supports the low density rib structures. 

The origin of the three-dimensional low density ribs within the cavities is still unclear and further research is needed to identify the causes. Previous studies of guide field magnetic reconnection simulations~\cite{Drake:2003} and a carried out spectral analysis of the wave in the cavity regions show that the frequency of the observed waves are in the range of the lower-hybrid frequency $\omega_{LH} = \omega_{pi}/\sqrt{1 + \omega_{pe}^2 / \Omega_{ce}^2}$, where $\omega_{pe}$ is the plasma frequency and $\Omega_{ce}$ the electron gyro-frequency. Spacecraft observations confirm the presence of lower hybrid waves also~\cite{Zhou:2011}. It has been suggested that lower-hybrid waves can be driven by electron beams~\cite{Drake:2003,McMillan:2007,Che:2009,Che:2010}, or by density gradients, as in the lower-hybrid drift instability~\cite{Scholer:2003}. The density ripples in the cavities have been previously detected in two dimensional Particle-in-Cell simulations, and spectral analysis suggest that the instabiity is related to the electron MHD Kelvin-Helmholtz mode \cite{Andrey:2011}. An additional possibility is that observed three-dimensional structures of higher density regions in proximity of the low density ribs are Alfv\'en vortex filaments~\cite{Alexandrova:2006,Xu:2010}, driven by drift kinetic Alfv\'en waves. The presence of Alfv\'en vortices have been observed in cusp region~\cite{Sundkvist;2005} and in proximity of a magnetic reconnection X lines~\cite{Chaston:2005, Chaston:2009}.

Moreover, the parallel electric field exhibits localized bipolar signatures in proximity of the cavity channels. These structures might be the signature of streaming instabilities such as the Buneman and/or two-stream instabilities~\cite{Andrey:2011}, or originate from the beam-driven lower hybrid instability~\cite{Che:2009}. A further study is required to identify the origins of the bipolar parallel electric field structures. It has been found that bipolar electric field structures are approximately located in the cavities between the low density ribs. One suggestive hypothesis is that the relative higher density regions in the cavities might act as waveguides, allowing the transport of bipolar electric field structures in channels between the low density ribs. 

\section{Conclusions}
This paper presented a simulation study of three dimensional structure of the cavity during collisionless magnetic reconnection with guide field.  This study shows the presence of further lower density regions within the cavities. These low density ribs are directed along the magnetic field lines and they are supported by an intense perpendicular electric field. Bipolar parallel electric field formations have been detected in the cavities between the low density ribs. Further research is necessary first to identify the instability or spectrum of instabilities leading to the low density rib structures and second to study their long-term evolution. In particular, the use of open boundary conditions in the outflow direction is necessary to avoid the plasma recirculation at the boundary, enabled by the periodic boundary of the presented simulation. The open boundary condition will allow to follow the evolution of the low density ribs over a large period of time. 
The existence of the low density ribs within the cavities, of perpendicular electric fields, and of bipolar electric field structures is very important the future NASA's Magnetospheric Multi Scale (MMS) mission. In particular, the reported features of guide field reconnection might be used to identify the proximity to the diffusion region. In fact, it has been suggested recently in Ref.~\cite{Lapenta:2011} that bipolar electric field signatures can be used to detect the diffusion region and to control the spacecraft towards magnetic reconnection site. This research suggests that additional signatures, such as the low density ribs and associated perpendicular electric fields, can possibly used as flag to detect the reconnection site also.

\begin{acknowledgments}
The present work is supported by NASA MMS Grant NNX08AO84G. Additional support for the KULeuven team is provided by the Onderzoekfonds KU Leuven (Research Fund KU Leuven) and by the European CommissionÕs Seventh Framework Programme (FP7/2007-2013) under the grant agreement no.~263340 (SWIFF project, www.swiff.eu). The simulations were conducted on the resources of the NASA Advanced Supercomputing Division (NAS).
\end{acknowledgments}

\providecommand{\noopsort}[1]{}\providecommand{\singleletter}[1]{#1}%

\end{document}